\documentclass{article}
 \pdfoutput=1
\usepackage{graphicx}  
\usepackage{amsmath}   
\usepackage{amssymb}   
\usepackage[export]{adjustbox}
\usepackage{bm} 
\usepackage{dcolumn}
\usepackage{color}
\usepackage{mathrsfs}
\usepackage{amsfonts}
\usepackage{varioref}
\usepackage{physics}
\usepackage[vcentermath]{youngtab}
\usepackage{wrapfig}
\RequirePackage[colorlinks,citecolor=blue,urlcolor=magenta,linkcolor=black]{hyperref}
\usepackage{tikz}
\usepackage{amsmath}
\usetikzlibrary{decorations.pathmorphing}
\usepackage[active]{srcltx}
\usepackage[numbers, sort&compress]{natbib}
\usepackage{tikz} 
\usepackage{subcaption}
\usepackage{comment}

\def\Tr{{\rm Tr}}

\labelformat{section}{Section #1} 
\labelformat{subsection}{Section #1} 
\labelformat{subsubsection}{Section #1}
\labelformat{subsubsubsection}{Section #1}
\labelformat{equation}{Eq.~(#1)} 
\numberwithin{equation}{section}
\labelformat{figure}{Fig.~#1} 
\labelformat{subfigure}{Fig.~\thefigure#1} 
\labelformat{table}{Tab.~#1} 
\labelformat{appendix}{Appendix #1}

\title{\bf {Replica Trick in Time-Dependent Geometries}}
\author{\bf Anastasios Irakleous$^{1}$\footnote{irakleous.anastasios@ucy.ac.cy}
\\
$^{1}$\small{Department of Physics, University of Cyprus, Nicosia 1678, Cyprus}
}

\date{ }  
\begin{document}
\maketitle

\begin{abstract}
We present a unified Lorentzian replica--path--integral framework for computing entanglement entropy in fully time-dependent quantum field theories and gravitational systems. Building on this framework, we apply the real-time replica trick to cosmological spacetimes without spatial boundaries and relate it to existing proposals that identify effective dual QFT descriptions on screen-like surfaces inside the spacetime. We also extend the Lorentzian replica construction to intrinsic gravitational subsystems, focusing on Hawking radiation and including quantum corrections in purely gravitational settings lacking time-reflection symmetry. In both contexts, island-type replica saddles arise naturally. A brief review of the QFT and holographic cases is included to establish the common structure underlying these applications.
\end{abstract}

\newpage
\tableofcontents


\section{Introduction}

The holographic principle, originally motivated by black hole thermodynamics
\cite{Bekenstein:1972tm,Hawking:1975vcx,tHooft:1993dmi,Susskind:1994vu}, has found a precise
realization in the AdS/CFT correspondence \cite{Maldacena:1997re,Gubser:1998bc,Witten:1998qj}.
This duality equates gravitational dynamics in a $(d+1)$-dimensional asymptotically Anti--de Sitter
(AdS) spacetime with a non-gravitational conformal field theory (CFT) living on its $d$-dimensional
boundary. Within this framework, entanglement entropy plays a central role in encoding emergent
geometry, through the Ryu--Takayanagi (RT) prescription \cite{Ryu:2006bv} and its covariant
generalization, the Hubeny--Rangamani--Takayanagi (HRT) proposal \cite{Hubeny:2007xt}. Boundary
subregions are associated with geometric duals in the bulk via extremal surfaces, which compute
entanglement entropy in both static and dynamical settings.

Quantum corrections to the RT and HRT prescriptions were formulated in
\cite{Faulkner:2013ana,Engelhardt:2014gca,Wall:2012uf,Freedman:2016zud,Headrick:2022nbe},
where it was shown that the entanglement entropy of a boundary region $A$ is given by the
\textit{generalized entropy}: the area of an extremal surface anchored to $\partial A$, plus the bulk
matter entropy across the region between the extremal surface and $A$. More recently, this framework
has been extended to include the \textit{island formula}
\cite{Penington:2019npb,Almheiri:2019psf,Almheiri:2019hni,Penington:2019kki,Almheiri:2019qdq,Page:1993wv,Marolf:2020rpm},
which allows for disconnected quantum extremal surfaces (QES). These developments have provided
important insights into the Page curve of evaporating black holes and the semiclassical restoration of
unitarity. For recent related work, see \cite{Nomura:2025whc,Ruan:2025uhl,Zhao:2025zgm,Rusalev:2025tdh}.

While AdS/CFT has been explored extensively, the extension of holographic principles to spacetimes
with positive cosmological constant, such as de Sitter (dS), remains less well understood
\cite{Gibbons:1977mu,Banks:2000fe,Witten:2001kn,Dyson:2002nt,Goheer:2002vf,Banks:2002wr,Banks:2003cg}.
Several proposals for dS holography have been put forward, including the dS/CFT correspondence
\cite{Strominger:2001pn,Bousso:2001mw} and analytic continuation from AdS setups, but a
first-principles nonperturbative formulation analogous to AdS/CFT is still lacking. This difficulty is
compounded by the absence of a spatial boundary on which to define a dual QFT. Nevertheless, recent
studies have explored analogs of entanglement entropy and quantum extremal surfaces in de Sitter
spacetime
\cite{Susskind:2021omt,Alishahiha:2004md,Alishahiha:2005dj,Dong:2010pm,Dong:2018cuv,Arias:2019pzy,Arias:2019zug,Arenas-Henriquez:2022pyh,Emparan:2022ijy,Svesko:2022txo,Panella:2023lsi}.
An interesting proposal identifies holographic screens with the pair of cosmological horizons associated
with antipodal comoving observers \cite{Susskind:2021omt}, leading to the \textit{bilayer proposal} for
holographic entanglement entropy
\cite{Susskind:2021esx,Shaghoulian:2021cef,Shaghoulian:2022fop} and its covariant extensions
\cite{Franken:2023pni,Franken:2023jas,Irakleous:2025trr}.

As in AdS/CFT, computing the entanglement entropy of a subregion $A$ in dS-like constructions
requires extremizing the generalized entropy of codimension-two surfaces homologous to $A$. However,
because the proposed dual theory is located \textit{inside} the spacetime rather than at its boundary,
there is an inherent ambiguity in identifying the allowed quantum extremal surfaces $\widetilde{A}$ and
in defining their associated entanglement entropy. A more fundamental definition of entropy therefore plays a particularly important role in these settings, since it offers a way to test which proposed constructions yield consistent and physically meaningful entanglement entropies.

A more fundamental formulation of entanglement entropy is provided by the replica trick and the
path-integral approach \cite{Calabrese:2004eu}. In quantum field theory, one computes the partition
function on an $n$-sheeted manifold obtained by gluing $n$ copies of the original spacetime along the
region $A$, and then analytically continues to $n\to 1$. When the QFT has a gravitational dual, this
construction lifts to a sum over bulk replica geometries subject to the corresponding boundary
conditions \cite{Lewkowycz:2013nqa,Dong:2016hjy}. In the semiclassical limit, the dominant replica
saddle reproduces the generalized entropy and yields the QES prescription. An advantage of the replica
method is that it remains applicable even when the dual theory resides inside the spacetime or when no
holographic dual is known.

Most applications of the replica trick assume Euclidean backgrounds, where branched covers and the
analytic continuation in $n$ are relatively straightforward. In general time-dependent semiclassical
geometries, however, the situation is considerably more subtle. Such geometries appear in quenches,
expanding spacetimes, black hole formation, and evaporation. Replica methods also apply directly to
gravitational theories \cite{Dong:2020iod}, where they provide a definition of
generalized entropy purely from the gravitational path integral. Cosmological spacetimes, such as de
Sitter, present additional challenges because Euclidean methods may need to be modified or replaced
entirely \cite{Maldacena:2012xp}, although replica-based entropy calculations exhibiting island
contributions have been carried out in specific low-dimensional de Sitter models
\cite{Balasubramanian:2020xqf}.

Covariant holographic constructions have been used in many time-dependent settings, including black
hole formation and evaporation \cite{Hubeny:2007xt,Akers:2019nfi}, de Sitter spacetime
\cite{Franken:2023pni}, closed FRW universes \cite{Franken:2023jas}, and bubble geometries
\cite{Irakleous:2025trr}. For general time-dependent states with a dual QFT on the boundary, the
classical covariant holographic entanglement entropy proposal has been derived from the Lorentzian
replica path integral in \cite{Dong:2016hjy}. Related real-time gravitational replica formalisms, including a variational principle for Lorentzian replica saddles and explicit low-dimensional examples, were developed in \cite{Colin-Ellerin:2020mva,Colin-Ellerin:2021jev}. In those works, geometric connections among replica copies are analyzed through a variational principle for Lorentzian saddles.

A natural question is whether the bulk replica path integral can be applied directly to gravitational
systems without a dual QFT. This would allow one to study the entanglement entropy of intrinsic
gravitational subregions---including Hawking radiation---without relying on an auxiliary boundary
theory. In simplified two-dimensional models such as JT gravity coupled to a CFT, the island formula has
been successfully implemented, with replica wormholes providing the dominant saddles responsible for
the emergence of islands and the reproduction of the Page curve
\cite{Penington:2019kki,Almheiri:2019qdq}. However, many existing implementations exploit
time-reflection symmetry and/or Euclidean replica constructions. To study more realistic scenarios, such
as black hole formation and evaporation in fully dynamical settings, and to address cosmological
geometries without spatial boundaries, it is useful to formulate the replica trick directly in Lorentzian
signature in a way that makes minimal reference to auxiliary boundary structures.
While related Lorentzian approaches exist, here we develop a more systematic replica–path–integral formulation for intrinsic gravitational subsystems without assuming any auxiliary dual description.

In this work, we develop a unified treatment of the replica trick for time-dependent geometries in four
contexts: (1) generic quantum field theories, (2) holographic QFTs with gravitational duals, (3)
cosmological spacetimes without spatial boundaries, and (4) purely gravitational theories without dual QFTs%
\footnote{Items (1) and (2) are included mainly for completeness and review, following the Lorentzian and
real-time replica constructions developed in \cite{Dong:2016hjy} and related work. Related foundational
work on real-time gravitational replica formalisms can be found in
\cite{Colin-Ellerin:2020mva,Colin-Ellerin:2021jev}.}.
Our aim is to provide a systematic Lorentzian replica-path-integral formulation of entanglement entropy
for gravitational subsystems in which no auxiliary boundary quantum field theory is available, and to
clarify how replica saddles, generalized entropy, and island-type contributions arise in cosmological
spacetimes without spatial boundaries. While real-time replica methods and replica wormholes have been
developed in holographic and low-dimensional settings, a unified treatment applicable to intrinsic
gravitational subsystems and screen-based cosmological constructions has remained less explicit. This
work fills that gap at the level of the semiclassical gravitational path integral.

For the first two cases, we review the Lorentzian replica construction developed in \cite{Dong:2016hjy}
and incorporate the replica wormhole contributions identified in \cite{Penington:2019kki,Almheiri:2019qdq}.
We then propose a generalization of this framework that allows for the computation of entanglement
entropy in cosmological spacetimes without boundaries, in a way that naturally interfaces with
screen-based descriptions such as the bilayer proposal, and that derives the generalized-entropy and
island structure directly from the Lorentzian replica construction rather than from phenomenological
extensions of boundary-based prescriptions.
Finally, we apply the bulk Lorentzian replica framework to extend the analysis of
\cite{Penington:2019kki,Almheiri:2019qdq} to the computation of entanglement entropy for intrinsic gravitational subsystems lacking time-reflection symmetry. For clarity, each case is presented in a largely
self-contained way.

The paper is organized as follows. In Section~2, we review the replica trick in non-gravitational quantum
field theories, emphasizing its formulation in real time and its applicability to time-dependent states.
Section~3 reviews the replica method in holographic theories with gravitational duals, including the
construction of the Lorentzian bulk replica path integral, the emergence of generalized entropy, and the
role of replica wormholes. Within this section, we also propose an extension of extremal surface
prescriptions to cosmological spacetimes without spatial boundaries. In Section~4, we apply the
Lorentzian replica path integral directly to gravitational systems without a dual QFT, focusing on the
computation of entanglement entropy for Hawking radiation in time-dependent geometries. We analyze
the structure of replica saddles and discuss the appearance of islands from replica wormholes. We also use a simplified toy model to illustrate the computation of entanglement entropy using only integer replica numbers and to include non-$\mathbb{Z}_n$-symmetric corrections. Such corrections have been shown to be important near phase transitions \cite{Penington:2019kki,Dong:2020iod,Akers:2020pmf}. We close with a discussion and conclusions in Section~5.

\section{Replica trick in non-gravitational theories}

The replica trick was first developed in quantum field theory by Calabrese and Cardy to compute the entanglement entropy of spatial subregions in static states, by evaluating partition functions on $n$-sheeted Riemann surfaces and analytically continuing in $n$ \cite{Calabrese:2004eu,Calabrese:2009qy}. In the gravitational context, a closely related construction underlies the derivation of the generalized entropy in time-independent spacetimes, where the replica method is performed in the Euclidean path integral and leads to the Ryu--Takayanagi prescription \cite{Ryu:2006bv,Lewkowycz:2013nqa}. In order to apply the replica trick to general time-dependent geometries, we need to formulate the replica path integral in Lorentzian signature. In this section, we review the construction of the Lorentzian replica path integral associated with the von Neumann entropy of a subregion of a QFT on a fixed curved spacetime. This construction will serve as the starting point for the gravitational generalizations discussed in later sections. We follow the derivation in \cite{Dong:2016hjy}.

We begin by choosing a subsystem $A$ of the QFT by splitting a Cauchy slice $S$ into two parts, $S=A\cup \bar{A}$, where $\bar{A}$ is the complement of $A$. The state on $A$ is described by the associated density matrix $\hat{\rho}_A$. The entanglement entropy of $A$ is given by
\begin{align}\label{S_A definition}
S_A = -\Tr(\hat{\rho}_A \log \hat{\rho}_A).
\end{align}
In general, it is difficult to compute the logarithm of the density matrix. Instead, we can use the relation
\begin{align}\label{eq:ent_entropy}
S_A = -\lim_{n \to 1} \frac{\log \Tr \hat{\rho}_A^n}{n-1}.
\end{align}
The calculation is reduced to finding an expression for $\Tr \hat{\rho}_A^n$ that can be analytically continued to non-integer values of $n$, at least near $n = 1$. This is the goal of the remainder of the section.

\subsection*{$\Tr\hat{\rho}_A^n$ as a path integral}

Consider a non-gravitational system in a pure state $\Psi$. This state can be specified by initial boundary conditions at $t=-\infty$ or by fixing the field configurations $\phi_s$ on a Cauchy slice $S'$ in the past of $S$, where $\phi_s$ denotes the fields of the theory. Consider also the wave functional $\Psi(\alpha)$ associated with a state $\ket{\alpha}$ on the Cauchy slice $S$, with $\phi_s=\phi^\alpha_s$ on $S$. We can express this wave functional as a path integral over all field configurations compatible with the state $\ket{\Psi}$ and the condition $\phi_s=\phi^\alpha_s$ on $S$,
\begin{align}
\Psi(\alpha) = \braket{\alpha}{\Psi}
= N^{-1/2} \int_\Psi^{\phi_s=\phi^\alpha_s {\ \rm on\ } S} [d\phi_s] e^{iS[\phi_s]},
\end{align}
where $N^{-1/2}$ is a normalization constant. The path integral is taken over all field configurations with finite action that satisfy the boundary conditions.

A pure state on a Cauchy slice is described by a density matrix of the form $\hat{\rho} = \ket{\Psi}\bra{\Psi}$. We can use a basis in which the matrix elements of $\hat{\rho}$ are given by two path integrals with boundary conditions on $S$,
\begin{gather}
\rho^{\alpha\alpha'} = \bra{\alpha} \hat{\rho} \ket{\alpha'} =
\braket{\alpha}{\Psi} \braket{\Psi}{\alpha'} = \Psi(\alpha) \Psi^*(\alpha').
\label{raa1}
\end{gather}
To obtain the corresponding matrix elements of the reduced density matrix associated with the subsystem $A$, we sum over the degrees of freedom of the subsystem $\bar{A}$. Using the notation $\ket{\alpha} = \ket{\alpha_A} \oplus \ket{\alpha_{\bar{A}}} \equiv \ket{\alpha_A}\ket{\alpha_{\bar{A}}}$ and the definition $\hat{\rho}_A = \Tr_{\bar{A}} \hat{\rho}$, we can write
\begin{align}
\rho_A^{\alpha_A \alpha'_A} &= \bra{\alpha_A} \hat{\rho}_A \ket{\alpha'_A}
= \bra{\alpha_A} \Tr_{\bar{A}} \big(\ket{\Psi} \bra{\Psi}\big) \ket{\alpha'_A} \nonumber \\
&= \bra{\alpha_A} \bigg( \sum_{\alpha_{\bar{A}}} \bra{\alpha_{\bar{A}}} \Psi \rangle \bra{\Psi} \ket{\alpha_{\bar{A}}} \bigg) \ket{\alpha'_A}
= \sum_{\alpha_{\bar{A}}} \braket{\alpha}{\Psi} \braket{\Psi}{\alpha'} \nonumber \\
&= \int [d\alpha_{\bar{A}}] \Psi(\alpha_A \oplus \alpha_{\bar{A}}) \Psi^*(\alpha_A' \oplus \alpha_{\bar{A}}) \nonumber \\
&= \int [d\alpha_{\bar{A}}]\left( N^{-\frac{1}{2}} \int_\Psi^{\phi_s=\phi^\alpha_s {\ \rm on\ } S} [ d\phi_s] e^{iS_{\uparrow}[\phi_s]}
N^{-\frac{1}{2}} \int^\Psi_{\phi_s=\phi^\alpha_s {\ \rm on\ } S} [ d\phi_s] e^{-iS_{\downarrow}[\phi_s]}\right) \nonumber \\
&= N^{-1} \int [d\phi_s] e^{iS[\phi_s]},
\label{rij1}
\end{align}
where in the fourth line we incorporate a Keldysh contour \cite{Keldysh:1964ud} that evolves $\ket{\Psi}$ forward in time and its conjugate backward. The sum over all configurations in the region $\bar{A}$ is equivalent to ``gluing''
\footnote{For time-independent states, this gluing provides a path integral over the original geometry from the past Cauchy slice to the time-reflection-symmetric Cauchy slice in the future. In time-dependent states without time-reflection symmetry around $S$, we still use the term ``glue'' to describe the identification of the ket and bra geometries along a region, but this braket geometry does not correspond to a physical spacetime geometry.}
the two path integrals in \ref{raa1} through the region $\bar{A}$. We obtain a geometry consisting of the original spacetime evolving forward in time up to a Cauchy slice $S=A\cup \bar{A}$ (the ``ket'' geometry), glued through the surface $\bar{A}$ to the same geometry evolving backward in time (the ``bra'' geometry). We denote this geometry by $\mathcal{R}_1$. The path integral \ref{rij1} is over all field configurations living on $\mathcal{R}_1$ that match the state $\ket{\Psi}$ and satisfy the boundary conditions $\phi_s = \phi^{\alpha_A}_s$ on the near ``past'' of $A$ and $\phi_s = \phi^{\alpha_A'}_s$ on the near ``future'' of $A$
\footnote{The terms past and future of $A$ are initially used in time-reflection-symmetric systems, where the geometry $\mathcal{R}_1$ can be realized as the full spacetime of the system. Here, we use ``past'' to refer to the boundary conditions on $A$ in the ket geometry and ``future'' to refer to the boundary conditions on $A$ in the bra geometry.}.

If the geometry is time-independent, the path integral is over configurations with future and past boundary conditions determined by $\Psi$ and a cut along the region $A$. For time-dependent geometries, we adopt this picture but must impose appropriate matching conditions on the near past ($A^-$) and near future ($A^+$) of the cut, to obtain a well-defined variational problem \cite{Dong:2016hjy}. The normalization factor $N^{-1}$ is equal to the path integral over ${\mathcal{R}}_1$ without the cut, ensuring that $\Tr \hat{\rho}_A = 1$.

Using the construction above, we can write the trace $\Tr\hat\rho_A^n$ as a path integral,
\begin{gather}\label{eq:trrn36}
\Tr\hat\rho_A^n=\sum_{i_1,i_2,\dots,i_n}\rho_A^{i_1 i_2}\rho_A^{i_2 i_3}\dots\rho_A^{i_n i_1}
=N^{-n}\int_{{\mathcal{R}}_n}[d\phi_s]e^{iS[\phi_s]},
\end{gather}
where the indices $i_m$ label states on the surface $A$ of the $m$th copy. Each matrix element $\rho_A^{ij}$ is a path integral of the form \ref{rij1}, and $\Tr\hat\rho_A^n$ involves $n$ copies of the geometry ${\mathcal{R}}_1$. The sum over $i_2$ is equivalent to gluing the near-future boundary of $A$ in the first copy to the near-past boundary of $A$ in the second copy. The same gluing applies to all indices $i_m$. The final sum (over $i_1$) glues the last copy to the first. We denote the resulting manifold by ${\mathcal{R}}_n$. Due to this gluing procedure, the full geometry on ${\mathcal{R}}_n$ contains conical singularities at the boundaries of $A$, where a cycle around $\partial A$ passes through all $n$ copies of $\mathcal{R}_1$. Moreover, the gluing along $A$ eliminates the dependence of the path integral on the bulk of $A$, leaving only a dependence on the boundary $\partial A$, as expected.

The path integral \ref{eq:trrn36} is all that is required to compute the entanglement entropy of the subsystem $A$. In principle, one first evaluates \ref{eq:trrn36} for arbitrary $n$ and then applies the result to \ref{eq:ent_entropy}. In practice, a full calculation of \ref{eq:trrn36} is possible only for very simple models. Typically, one assumes that the dominant contributions to the path integral arise from $\mathcal{R}_n$ geometries with $\mathbb{Z}_n$ symmetry, for which the field configurations on the $n$ copies of $\mathcal{R}_1$ are identical \footnote{Throughout this work we assume semiclassical dominance of replica-symmetric saddles. The 
existence and stability of such saddles in fully general time-dependent geometries is not proven and may 
depend on the details of the spacetime and contour prescription. Our results should therefore be understood 
as conditional on this assumption.}
. To further simplify the analysis, one focuses on the behavior of \ref{eq:trrn36} in the vicinity of $n=1$.

\section{Replica method in gravitational theories dual to QFTs}

\subsection{Setting up the bulk replica path integral}\label{Bulk RPI}

Via holography, we can extend the replica calculation of the previous section to theories with dynamical
geometry, where the gravitational path integral is anchored by boundary conditions supplied by a dual
QFT. Suppose that the $(D-1)$-dimensional QFT discussed above is dual to a $D$-dimensional
gravitational ``bulk'' theory. We assumed that the QFT lives on a fixed background geometry; such a
geometry is naturally defined in the asymptotic region of a dynamical spacetime. For the two theories to
be dual, the partition function of the QFT must coincide with that of the bulk theory. To define the
partition function, we perform a path integral over all field configurations allowed by the boundary
conditions associated with the state of the system as specified in the dual QFT. For the bulk
gravitational theory, we must also sum over all geometries and topologies compatible with the same
boundary conditions.

When the QFT lives on the boundary of the bulk spacetime, holography implies that we can compute the
path integral \ref{eq:trrn36} using a bulk path integral over all geometries $\mathcal{M}_n$ with
boundary $\mathcal{R}_n$,
\begin{align}\label{dM=R}
    \partial \mathcal{M}_n=\mathcal{R}_n.
\end{align}
Each wave functional appearing in \ref{eq:trrn36} can be represented by a bulk path integral over
geometries whose boundary coincides with the QFT geometry. By gluing the $2n$ bulk geometries, we
construct the manifold $\mathcal{M}_n$.

The geometry $\mathcal{M}_1$ is obtained by gluing a ``ket'' and a ``bra'' geometry. The ket geometry
has initial boundary conditions determined by the state of the system and future boundary conditions
specified on a Cauchy slice $\Sigma_S$. Similar to the QFT case, we fix the metric and field
configurations on $\Sigma_S$. We take both the ket and bra bulk geometries to have future boundary
$\Sigma_S$, with $\partial\Sigma_S=S$, where $S=A\cup \bar{A}$. Note that a given $S$ does not uniquely
determine a bulk Cauchy slice $\Sigma_S$. Any Cauchy slice $\Sigma_S'$ satisfying $\partial\Sigma_S'=S$
is allowed. Following \cite{Dong:2016hjy}, we sum over all such admissible $\Sigma_S'$. This sum is
required in order to include in the bulk path integral all geometries $\mathcal{M}_n$ compatible with
$\mathcal{R}_n$.

Moreover, since the entanglement entropy is the same for any non-timelike surface $A'$ with
$\partial A'=\partial A$, we may use any such $A'$ to compute $S_A$. This corresponds to choosing
different bulk Cauchy slices. As we will see, the extremal surface $\tilde{A}$ relevant for the
computation of $S_A$ and the generalized entropy depends only on the endpoints of $A$, which are the
same for every $A'$. Because different $A'$ correspond to the same physical quantity, we should not
include in the path integral a sum over all $A'$ and the associated bulk Cauchy slices. Therefore, we
restrict the allowed bulk Cauchy slices by imposing the condition $\partial\Sigma_S'=S$.

When the gravitational action on $\mathcal{M}_n$ is large, as in the case of a large black hole or a
cosmological spacetime, we can use the saddle-point approximation for the bulk path integral. In this
approximation, the geometry $\mathcal{M}_n$ is fixed by the classical equations of motion. Leading
quantum corrections can be incorporated by including in the equations of motion the expectation value
of the energy-momentum tensor of the matter and metric fields. This is commonly referred to as the
\textit{semiclassical approximation}.

Since the boundary conditions defining $\mathcal{M}_n$ possess $\mathbb{Z}_n$ symmetry (both in the
initial conditions and in the geometry $\mathcal{R}_n$), we assume that the dominant semiclassical bulk
saddle also has $\mathbb{Z}_n$ symmetry. We further restrict attention to geometries $\mathcal{M}_n$
that are smooth everywhere except at $\partial A$, where conical singularities with total opening angle
$2\pi n$ appear. Under these assumptions, the geometry $\mathcal{M}_n$ can be decomposed into $n$
identical copies, denoted $\tilde{\mathcal{M}}_1$. $\tilde{\mathcal{M}}_1$ is the fundamental domain of
the $\mathbb{Z}_n$ orbifold. It is not the same as $\mathcal{M}_1$: if we glue $n$ copies of
$\mathcal{M}_1$ together, we obtain conical singularities with opening angle $2\pi n$ at the fixed
points of the $\mathbb{Z}_n$ symmetry, similar to those appearing at $\partial A$ in $\mathcal{R}_n$.

The copies $\mathcal{R}_1$ of the geometry $\mathcal{R}_n$ are continuous along the surface $\bar{A}$
and are glued cyclically along $A$. Requiring $\partial\mathcal{M}_n=\mathcal{R}_n$, the copies
$\tilde{\mathcal{M}}_1$ must be glued in the same way as the copies of $\mathcal{R}_1$. The bulk Cauchy
slice is split along a surface $\tilde{A}$, $\Sigma_S=\Sigma_A\cup\Sigma_{\bar{A}}$, such that
$\partial\Sigma_A=A\cup\tilde{A}$ and $\partial\Sigma_{\bar{A}}=\bar{A}\cup\tilde{A}$. The copies
$\tilde{\mathcal{M}}_1$ are continuous along $\Sigma_{\bar{A}}$ and are glued cyclically along
$\Sigma_A$. The key difference with the construction of $\mathcal{R}_n$ is that the geometry around
$\tilde{A}$ is smooth, which, as we will see, leads to the area (gravitational) contribution to the
entanglement entropy. As in the QFT case, by ``gluing'' we mean that we fix the metric and field
configurations on $\bar{A}$ for the ket and bra geometries and then sum over all allowed such
configurations (in situations with time-reflection symmetry across $\Sigma_S$, this bra--ket geometry
coincides with the full spacetime of the bulk theory).

\subsection*{Bulk path integral}

According to holography, the replica path integral of the QFT is equivalent to a bulk path integral. We
will call this path integral the \textit{bulk replica path integral}, and we assume it is of the form
\begin{gather}
G_n(A) = \int [dgd\phi_s] e^{iS[g,\phi_s]}.
\label{eq:gravitational path integral}
\end{gather}
We impose boundary conditions compatible with the state $\Psi$ and include geometries $\mathcal{M}_n$
satisfying \ref{dM=R}. The $\Tr\hat{\rho}_A^n$ is given by
\begin{gather}
\Tr\hat{\rho}_A^n = \frac{Z_n(A)}{Z_1^n}= \frac{G_n(A)}{G_1^n},
\label{san}
\end{gather}
where $Z_n(A)$ is the unnormalized path integral in \ref{eq:trrn36}. We can now express
\ref{eq:ent_entropy} using only the bulk theory,
\begin{gather}
S_A = -\lim_{n \to 1} \frac{\log G_n(A) - n \log G_1}{n-1}.
\label{svn}
\end{gather}

For gravitational theories coupled to matter fields, we can split the action into two parts: the
``gravitational'' part and the matter part,
\begin{align}
    S[g,\phi_g,\phi_m] = S_{\rm grav}[g,\phi_g] + S_{\rm matter}[g,\phi_g,\phi_m],
\end{align}
where $\phi_m$ denotes the matter fields of the theory and $\phi_g$ denotes the remaining fields, such
as the dilaton in two-dimensional theories. We can integrate out the matter fields,
\begin{align}
    G_n(A) = \int [dgd\phi_g] e^{iS_{\rm grav}[g,\phi_g]} \int [d\phi_m] e^{iS_{\rm matter}[g,\phi_g,\phi_m]}
    \equiv \int [dgd\phi_g] e^{iS_{\rm grav}[g,\phi_g] + \log G_{\rm matter}[g,\phi_g]}.
    \label{eq:45}
\end{align}
The dominant contributions to \ref{eq:45} come from saddle points, where $g$ and $\phi_g$ satisfy the
equations of motion associated with the exponent $iS_{\rm grav}[g,\phi_g] + \log G_{\rm matter}[g,\phi_g]$%
\footnote{This saddle-point treatment of Lorentzian replica path integrals is closely related to the
real-time gravitational replica framework developed in \cite{Colin-Ellerin:2020mva,Colin-Ellerin:2021jev},
where Lorentzian replica saddles are characterized through a variational principle, primarily as geometric connections among replica copies.}.
We can think of the first term as the classical part of the effective action, and the second term as the
quantum corrections from the matter fields%
\footnote{There are also quantum corrections from the metric and $\phi_g$ fields.}.
The path integral takes the form%
\footnote{In general there can be more than one saddle, but typically one saddle dominates due to the
large value of the exponent.}
\begin{align}\label{36}
    G_n(A) \sim e^{iS_{\rm grav}[g_c,\phi_{g,c}](\mathcal{M}_n) + \log G_{\rm matter}[g_c,\phi_{g,c}](\mathcal{M}_n)},
\end{align}
where $g_c$ and $\phi_{g,c}$ solve the equations of motion associated with the exponent%
\footnote{We absorb the first-order quantum contributions from $g$ and $\phi_g$ into
$\log G_{\rm matter}[g_c,\phi_{g,c}](\mathcal{M}_n)$.}. We explicitly indicate the dependence on
$\mathcal{M}_n$ for later reference. Applying the above relation to \ref{svn}, we obtain
\begin{align}\label{4.7}
    S_A = &-\lim_{n\to1} \frac{iS_{\rm grav}[g_c,\phi_{g,c}](\mathcal{M}_n) - n iS_{\rm grav}[g_c,\phi_{g,c}](\mathcal{M}_1)}{n-1}
    \nonumber\\
    &-\lim_{n\to1} \frac{\log G_{\rm matter}[g_c,\phi_{g,c}](\mathcal{M}_n) - n \log G_{\rm matter}[g_c,\phi_{g,c}](\mathcal{M}_1)}{n-1}.
\end{align}
Note that in the absence of gravity (a QFT defined on a fixed geometry) only the second term appears.
We can use the second term to define the \textit{matter entropy} associated with $\Sigma_A$, i.e., the
entanglement entropy of the matter fields on $\Sigma_A$%
\footnote{When gravity is dynamical, the ``matter entropy'' also includes contributions from the $g$ and
$\phi_g$ degrees of freedom.}. This picture is meaningful in the semiclassical approximation, where the
full quantum system can be approximated by matter and metric fields fluctuating around a fixed
background geometry.

\subsection{Generalized entropy}\label{s gen section}

According to the first line in \ref{4.7}, we need to compute the difference between the gravitational
action on ${\mathcal{M}}_n$ and $n{\mathcal{M}}_1$ in the limit $n \to 1$. For $n = 1$, the geometry on
${\mathcal{M}}_n$ coincides with the original geometry $\mathcal{M}_1$, so the difference arises only at
order $n - 1$. As argued earlier, the $n$ copies are glued together along the region $\Sigma_A$, and the
surface $\tilde{A}$ is the fixed point of the $\mathbb{Z}_n$ symmetry.

It is convenient to introduce another geometry, composed of $n$ copies of $\mathcal{M}_1$ glued
together via $\Sigma_A$%
\footnote{Although $\Sigma_A$ depends on $n$ through the equations of motion, in the limit
$n\rightarrow 1$ it can be treated as independent of $n$.}. We denote this geometry by
$\tilde{\mathcal{M}}_n$. $\tilde{\mathcal{M}}_n$ has a conical singularity along $\tilde{A}$ with total
opening angle $2\pi n$ (corresponding to a deficit angle $2\pi(1-n)$). The gravitational action on
$\tilde{\mathcal{M}}_n$ is simply $n$ times that on $\mathcal{M}_1$%
\footnote{No contribution from the real conical singularity is included in this evaluation.},
so the first limit in \ref{4.7} can be expressed as the difference between the actions evaluated on
$\tilde{\mathcal{M}}_n$ and $\mathcal{M}_n$.

We can write the action on ${\mathcal{M}}_n$ for $n \to 1$ as%
\footnote{The term $nS_1$ arises because the integral extends over $n$ times the domain of the $n=1$
integral.}
\begin{gather}
S_n\Big|_{n\rightarrow1} = \left(nS_1 + (n-1)\int d^Dx \, \partial_n L[\rho,\phi_g]\right)_{n\rightarrow1},
\end{gather}
where $S_{\rm grav}(\mathcal{M}_n) \equiv S_n$ and we have dropped indices from $g_c$ and $\phi_{g,c}$
for simplicity. The term $nS_1$ corresponds to the action integrated over the singular geometry, without
contributions from the singularity. Away from the $\mathbb{Z}_n$-fixed points, we have
\begin{gather}
\int d^Dx \, \partial_n L[\rho,\phi] = \int d^Dx \left( \frac{\delta L}{\delta \rho} \partial_n (\delta \rho)
+ \frac{\delta L}{\delta \phi} \partial_n (\delta \phi) \right).
\end{gather}
The variation of the action with respect to the fields vanishes at leading order in $(n-1)$ due to the
equations of motion. Thus, the only non-vanishing contribution to $S_n - nS_1$ comes from the
neighborhood of the $\mathbb{Z}_n$-fixed points. Near these points, ${\mathcal{M}}_n$ is regular and
therefore differs from the singular geometry $\tilde{\mathcal{M}}_n$. At leading order in $(n-1)$, we
can therefore replace $\mathcal{M}_n$ by $\tilde{\mathcal{M}}_n$ with the conical singularity
appropriately regularized.

Near a $\mathbb{Z}_n$ fixed point, the metric can be written as
\begin{gather}
ds^2 = -r^2 dt^2 + h(r) dr^2 + ds_{D-2}^2 = r^2 d\tau^2 + n^2 dr^2 + ds_{D-2}^2 + \mathcal{O}\big((n-1)\big),
\end{gather}
where the singularity is at $r=0$ and $ds_{D-2}^2$ is the transverse part of the geometry. Here
$\tau = it$ is a local coordinate near the fixed point, not a Wick rotation%
\footnote{We choose locally flat coordinates near the $\mathbb{Z}_n$-fixed point for simplicity.}. The
fields are periodic under $\tau \to \tau + 2\pi n$, and completing a full cycle requires passing through
all $n$ copies. The function $h(r)$ satisfies $h(r) = n^2 + \mathcal{O}(r^2)$ near the singularity and
$h(r) = 1$ away from it%
\footnote{A convenient choice is $h(r) = 1 + (n-1)e^{r^2/a^2}$ \cite{Lewkowycz:2013nqa}, which
explicitly shows that $\mathcal{M}_n$ differs from $\mathcal{M}_1$ by order $(n-1)$.}. It serves as a
regulator for the geometry near the conical singularity.

The contribution from the regularized tip of the cone to the gravitational action is%
\footnote{A more careful discussion about the evaluation of this contribution in Lorentzian signature can
be found in \cite{Dong:2016hjy,Colin-Ellerin:2020mva,Colin-Ellerin:2021jev}.}
\begin{align}\label{eq:cone_contribution}
i \int dtdrd^{D-2}x\sqrt{-g}R = \int d\tau drd^{D-2}x \sqrt{g} R
= 4\pi(1-n) \textrm{Area}(\tilde{A}),
\end{align}
where $\textrm{Area}(\tilde{A})$ is the area of the codimension-2 surface $\tilde{A}$
\cite{Lewkowycz:2013nqa}. Equivalently, this term can be viewed as arising from the localized curvature
at the (regularized) tip. Note that this
contribution is real \cite{Colin-Ellerin:2020mva,Colin-Ellerin:2021jev}. For exact Lorentzian replica solutions in JT gravity, this factor was also derived
in \cite{Blommaert:2023vbz}.

In two-dimensional theories, the action typically includes a term
\(\frac{1}{16\pi G_N} \int dt \, dr \, R F(\phi)\), where \(F(\phi)\) is a finite function of the dilaton.
The contribution from a $\mathbb{Z}_n$ fixed point is
\begin{gather}\label{eq:cone_contribution_2D}
i \int dt \, dr \, \sqrt{-g} R F(\phi) = 4\pi (1-n) F(\phi(\tilde{A})).
\end{gather}

Since $\mathcal{M}_n$ coincides with $\tilde{\mathcal{M}}_1$ away from the fixed points,
\ref{eq:cone_contribution} gives
\begin{align}\label{eq:35}
iS_{\rm grav}[g_c,\phi_{g,c}](\mathcal{M}_n) =
n\, iS_{\rm grav}[g_c,\phi_{g,c}](\mathcal{M}_1) + \frac{1}{16\pi G}\, 4\pi(1-n)\, \textrm{Area}(\tilde{A}).
\end{align}

In the dominant contribution to the replica path integral \ref{eq:45}, the replicas are glued via the
region $\Sigma_A$. In the limit $n \to 1$, the geometry $\mathcal{M}_n$ coincides with the geometry
constructed by $n$ copies of $\mathcal{M}_1$ glued cyclically via $\Sigma_A$, which is also used to
define the matter entropy of the fields on $\Sigma_A$. Thus, the second limit in \ref{4.7} corresponds
to the entropy of the matter fields in $\Sigma_A$. Applying \ref{eq:35} for the first term in \ref{4.7}
and replacing the second term with the matter entropy, we obtain the generalized entropy
\begin{align}\label{s gen 3.14}
S_A = \frac{\textrm{Area}(\tilde{A})}{4G} + S_{\rm matter}(\Sigma_A).
\end{align}

The location of $\tilde{A}$ is dynamical; we must sum over all allowed configurations in the path
integral. The dominant contribution comes from the $\tilde{A}$ that extremizes the generalized entropy.
If there are multiple extrema, the dominant contribution is from the configuration with minimal
generalized entropy, since $G_n(A) \sim e^{-(n-1)S_{\rm gen}} G_1^n$.

\subsection{Replica wormholes}

Using the replica construction of the previous section, we can naturally include \textit{islands} in the
calculation. This has been done in a similar context for gravitational theories without a dual
description, leading to the island rule and its corrections for time-independent semiclassical
geometries \cite{Lewkowycz:2013nqa,Penington:2019kki,Almheiri:2019qdq}.

The island-rule entropy appears naturally when we allow semiclassical geometries $\mathcal{M}_n$ in
which the $n$ copies $\tilde{\mathcal{M}}_1$ are glued cyclically not only through the surface
$\Sigma_{A}$, but also through an additional surface $I$. Such geometries are not excluded by the
boundary conditions imposed by the geometry $\mathcal{R}_n$, and in principle they should be included
in the gravitational path integral%
\footnote{For a discussion of factorization issues arising in such geometries, see
\cite{Penington:2019kki,Giddings:2020yes}.}. We refer to these geometries as \textit{replica wormholes}%
\footnote{Here we use the term ``replica wormholes'' to denote saddle geometries in which replica copies
are connected in a way that gives rise to island-type contributions to entanglement entropy. This usage
differs from that in \cite{Colin-Ellerin:2020mva,Colin-Ellerin:2021jev}, where Lorentzian replica wormholes refer to the connections between the copies only via $\Sigma_A$.}.
We note that the path integral should also include replica wormholes without $\mathbb{Z}_n$ symmetry
and configurations in which the $n$ copies are not cyclically connected, although such contributions
are expected to be exponentially suppressed by a characteristic entropy of the system.

As in the previous section, we assume a QFT in a state $\Psi$ defined on a fixed geometry with a dual
bulk gravitational description. We focus on states in which the bulk geometry admits a semiclassical
description. To compute the entanglement entropy of a subregion $A$ of the QFT, we consider a path
integral on the geometry $\mathcal{R}_n$ over all field configurations compatible with the boundary
conditions determined by the state $\Psi$. This path integral admits a gravitational dual description
as a path integral over all bulk field configurations compatible with the dual boundary conditions and
over all geometries $\mathcal{M}_n$ satisfying $\partial\mathcal{M}_n=\mathcal{R}_n$. Assuming
$\mathbb{Z}_n$ symmetry and working near the limit $n\rightarrow 1$, the geometry $\mathcal{M}_n$
approaches, at leading order in $(n-1)$, a configuration consisting of $n$ copies of $\mathcal{M}_1$
except at the codimension-2 fixed points of the $\mathbb{Z}_n$ symmetry, $\tilde{A}$, homologous to $A$,
where a regulated conical singularity with opening angle $2\pi n$ appears%
\footnote{For non-integer $n$ there are no literal $n$ copies glued together; the analytic continuation to
$n\to1$ applies to the value of the action evaluated on $\mathcal{M}_n$. We use this picture because it
provides an intuitive way to visualize the role of islands. Equivalently, one may study the quotient
geometry $\mathcal{M}_n/\mathbb{Z}_n$.}. The $n$ copies are glued together along a codimension-1
surface $\Sigma_A$ with $\partial\Sigma_A=A\cup\tilde{A}$. This is the construction used to obtain
\ref{s gen 3.14}.

To obtain $\mathbb{Z}_n$-symmetric replica wormholes, we now allow the $n$ copies to be glued together
along an additional non-timelike codimension-1 surface $I$. This surface may be disconnected. The
boundaries of $I$ are fixed surfaces of the $\mathbb{Z}_n$ symmetry. Because $\mathcal{M}_n$ is
constructed using a bulk Keldysh contour, the surface $I$ must lie on the same bulk Cauchy slice as
$\Sigma_A$%
\footnote{The ``ket'' and ``bra'' bulk geometries are glued along a Cauchy slice on which future boundary
conditions are imposed.}.

As for $\tilde{A}$, in the limit $n\rightarrow1$ a regularized conical singularity appears near
$\partial I$, contributing a surface term analogous to \ref{eq:35}. Moreover, since the $n$ bulk copies
are glued together along $\Sigma_A\cup I$, the matter partition function in \ref{4.7} computes the
semiclassical entropy of the matter fields on $\Sigma_A\cup I$. Using these facts, \ref{4.7} becomes
\begin{align}\label{s gen with I}
    S_A = \frac{\textrm{Area}(\tilde{A}\cup \partial I)}{4G} + S_{\rm matter}(\Sigma_A\cup I).
\end{align}

Beyond providing the entropy of $A$, the semiclassical nature of the replica path integral leads to
practical rules for determining the island $I$ and its contribution. First, as noted above, $I$ must
lie on the same bulk Cauchy slice as $\Sigma_A$. Second, the location of $I$ is a dynamical variable in
the path integral, and the dominant contributions arise from configurations in which $I$ extremizes
the action, or equivalently the generalized entropy. If multiple such configurations exist, including
the no-island configuration, the one with minimal generalized entropy dominates. Finally, $I$ may be
disconnected, corresponding to the presence of multiple islands.

The inclusion of replica wormholes and islands can thus be viewed as a generalization of
\ref{s gen 3.14}, in which $\tilde{A}$ and $\Sigma_A$ may themselves be disconnected. Such
configurations are classically forbidden, since if $\tilde{A}\cup \partial I$ is extremal then
$\tilde{A}$ is also extremal%
\footnote{A configuration is extremal if any infinitesimal variation of the shape of
$\tilde{A}\cup \partial I$ leaves the action unchanged, including variations that keep $\partial I$
fixed.}
and the area $\textrm{Area}(\tilde{A}\cup \partial I)$ is strictly larger than
$\textrm{Area}(\tilde{A})$.

\subsection{Cosmologies without boundaries}

In the previous sections we reviewed the basic ideas needed to apply the replica trick in gravitational theories with a timelike boundary, where one can naturally define a non-gravitational QFT dual to the bulk theory. 
Cosmologies such as de Sitter lack such a boundary. 
In such cases, some proposals associate a putative dual QFT with a non-spacelike codimension-1 surface in the interior of the bulk spacetime, rather than with an asymptotic boundary.
Efforts to propose a consistent holographic description in de Sitter spacetime have led,
among others, to the \textit{bilayer proposal}, which yields sensible results for entanglement entropies
of subregions and their related bulk domains \cite{Franken:2023pni}. A natural generalization of this
proposal has also been applied to FRW cosmologies \cite{Franken:2023jas} and bubble geometries
\cite{Irakleous:2025trr}. Despite its success, the bilayer proposal is, at present, a phenomenological
extension of the HRT prescription. Moreover, it is not a priori clear how to incorporate quantum
corrections systematically, since the classical bilayer prescription can involve disconnected area
contributions. The goal of this subsection is to show how a Lorentzian replica-path-integral
construction can be formulated in spacetimes without spatial boundaries and how it reproduces the
generalized-entropy/island structure, including the leading semiclassical (one-loop) corrections, once a
screen-based dual description is assumed.

The first step is to choose a non-spacelike codimension-1 bulk surface on which to place the QFT; we
call it the \textit{screen trajectory}%
\footnote{It would be interesting to use replica-path-integral reasoning to constrain, or help identify,
the correct dual description in cosmological spacetimes without spatial boundaries. In this work,
however, we take existing screen-based dual proposals as a working framework and study how the
Lorentzian replica path integral and replica-wormhole (island) contributions are realized within this
setting.}.
The screen trajectory is treated as fixed, allowing only small quantum fluctuations of the metric that can be described perturbatively around a semiclassical background.
On this fixed background one defines a QFT living on the screen trajectory. Note that this QFT is not part of the bulk gravitational theory; rather, the bulk geometry is used to specify a candidate non-gravitational theory that may be dual to (or capture some sector of) the bulk dynamics.
This step is meaningful in bulk theories admitting semiclassical codimension-1 non-spacelike surfaces,
such as de Sitter spacetime and FRW cosmologies where static patch holography has been applied. Any
bulk Cauchy slice intersects the screen trajectory along a Cauchy slice of the QFT, which we call the
\textit{screen}. A special case of this construction is when the screen trajectory coincides with a
spatial bulk boundary, in which case we recover the setup of the previous sections.

This criterion alone does not isolate a unique screen trajectory, since one can define a QFT on many
non-spacelike slices of the bulk geometry. Moreover, not every screen trajectory can plausibly support a
dual description of the \emph{entire} bulk spacetime, since a holographic description is expected to be
limited by an entropy bound controlled by the area of appropriate screens in Planck units. In the cases
where the bilayer proposal has been studied \cite{Franken:2023pni,Franken:2023jas}, the authors consider
screen trajectories associated with two comoving observers in the bulk geometry and use the Bousso
covariant entropy bound \cite{Bousso:1999xy,Bousso:2002ju} to identify the bulk region that such a QFT
can ``trivially'' encode. The bilayer proposal applied to these screen trajectories suggests that the
QFT can encode a larger bulk region than this bound would indicate. This is true for a
distinguished screen trajectory in de Sitter \cite{Franken:2023pni}, and for a family of screen
trajectories in some FRW cosmologies \cite{Franken:2023jas}. For other FRW cosmologies, the bilayer
proposal does not yield a QFT with enough degrees of freedom to describe the full bulk geometry, but
rather only a subset of it.

After choosing a screen trajectory, we can apply the replica trick for a QFT living on this trajectory.
First, we choose a subregion $A$ of a QFT Cauchy slice $S$. The goal is to calculate the entanglement
entropy of $A$ with its complement $\bar{A}$, where $S=A\cup \bar{A}$. To create the replica manifold we start from
some initial state on $S'$, in the past of $S$, and evolve in real time up to $S$. We then ``glue'' this
geometry along $\bar{A}$ to the same geometry evolved back in time from $S$ to $S'$ to form $\mathcal{R}_1$.
Next, we glue $n$ copies of $\mathcal{R}_1$ cyclically along $A$ (as described in the previous section)
to form the replica geometry $\mathcal{R}_n$. Calculating the replica path integral $Z_n$ on
$\mathcal{R}_n$ and taking the limit $n\rightarrow1$ yields the entanglement entropy of $A$ via
\ref{san} and \ref{eq:ent_entropy}.

In the spirit of the holographic principle, we assume that the replica path integral on $\mathcal{R}_n$
is captured by a bulk replica path integral $G_n$ involving bulk geometries $\mathcal{M}_n$. If the dual
theory is located on a spatial boundary of the bulk, one imposes $\partial\mathcal{M}_n=\mathcal{R}_n$.
In the absence of such a boundary, a natural generalization is to require that the replica geometry
$\mathcal{R}_n$ is embedded in the bulk geometry,
\begin{align}
\mathcal{R}_n\subset\mathcal{M}_n.
\label{Rn in Mn}
\end{align}
Accordingly, in the bulk replica path integral we include bra--ket bulk geometries compatible with the
state $\Psi$ that contain an embedded copy of $\mathcal{R}_n$, and we sum over bulk fields and
geometries subject to this condition. We also assume that the dominant semiclassical contributions come
from bulk saddles that are smooth away from the conical singularities associated with $\partial A$%
\footnote{The condition \ref{Rn in Mn} should be viewed as a working hypothesis for implementing the
replica boundary conditions in a spacetime without a spatial boundary. It is motivated by the boundary
case $\partial\mathcal{M}_n=\mathcal{R}_n$ and by the expectation that the QFT replica contour specifies
an interior hypersurface in the bulk path integral. A more intrinsic characterization of the allowed
bulk boundary conditions in cosmology would be valuable, but is beyond the scope of this work.}.
Note that the condition \ref{Rn in Mn} can also be used for geometries with boundaries, but in many
studied cases a QFT located on the boundary of the spacetime is expected to have enough degrees of
freedom to encode the full bulk region.

Near the limit $n\rightarrow1$ the replica path integral has properties analogous to the boundary case.
To see this, it is useful to think of $\mathcal{M}_n$ as built from $n$ copies of a bra--ket geometry
$\tilde{\mathcal{M}}_1$%
\footnote{We assume $\mathbb{Z}_n$ symmetry.}, where the ket and bra parts are obtained from the
evolution of an initial state on the Cauchy slice $\Sigma'$ up to some Cauchy slice $\Sigma$. The
condition $\mathcal{R}_n\subset\mathcal{M}_n$ can then be written as $\mathcal{R}_1\subset\tilde{\mathcal{M}}_1$.
Because the copies $\mathcal{R}_1$ are glued along $\bar{A}$ in $\mathcal{R}_n$, the copies
$\tilde{\mathcal{M}}_1$ should be glued along a bulk codimension-1 surface ${\Sigma}_{\bar{A}}$, with
$\bar{A}\subset{\Sigma}_{\bar{A}}$. Similarly, the copies should be glued cyclically along $\Sigma_A$,
with $A\subset\Sigma_A$. These two surfaces form a bulk Cauchy slice: $\Sigma_A\cup{\Sigma}_{\bar{A}}=\Sigma$.

As in the previous sections, we include all geometries $\mathcal{M}_n$ compatible with the state $\Psi$
and the replica data specified by $\mathcal{R}_n$. Assuming replica symmetry for $\mathcal{M}_n$ in the
semiclassical approximation and working near the limit $n\rightarrow1$, the dominant contributions come
from geometries $\mathcal{M}_n$ with a regularized conical defect around a codimension-2 fixed surface
$\tilde{A}$, homologous to $A$. The contribution to the action from the regularized defect is
${\rm Area}(\tilde{A})/(4G\hbar)$. Note that in the present setting the conical singularities at
$\partial A$ lie in the interior of the spacetime rather than on its boundary. Similar to the Euclidean
derivation of \cite{Lewkowycz:2013nqa}, we do not include independent gravitational boundary or defect
terms localized at these singularities. The validity of this prescription in fully Lorentzian settings
merits further investigation.

An advantage of the replica method is that it incorporates quantum corrections directly through the
replica path integral. The second term in \ref{4.7} has the same form as before. Since the $n$ copies of
$\mathcal{M}_n$ are glued together via the region $\Sigma_A$, the matter functional in \ref{4.7}
computes the semiclassical entropy of quantum fields on $\Sigma_A$. Together with the area term, this
yields the generalized entropy.

In the bulk path integral we sum over all possible choices of $\Sigma$, with $S\subset\Sigma$, and all
possible choices of $\Sigma_A$, with $A\subset\Sigma_A$ and $\bar{A}\cap\Sigma_A=\emptyset$. The
dominant contributions come from saddle points of the action, which correspond to extremizing the
generalized entropy with respect to the position of $\tilde{A}$. If there is more than one such saddle,
the dominant contribution comes from the one with minimum action.

The above treatment can be applied directly to de Sitter spacetime. Following the bilayer proposal, we
define the QFT on the two horizons associated with a pair of comoving observers, as described in
\cite{Franken:2023pni}. The QFT Cauchy slice $S$ is the union of two disjoint screens, each associated
with one horizon, and a bulk Cauchy slice $\Sigma$ with $S\subset\Sigma$ is split into three parts. A
general choice of $A$ involves both screens and $\tilde{A}$ can be disconnected. The replica
construction then implies that the entanglement entropy of $A$ with its complement is given by the
generalized entropy \ref{s gen 3.14}, and if $\Sigma_A$ is disconnected the generalized entropy takes
the form \ref{s gen with I}. This is precisely the generalized-entropy structure used in
\cite{Franken:2023pni}. Furthermore, the generalized entropy should be extremized over the position of
$\tilde{A}$, and if more than one extremum exists, the dominant contribution is the one with minimum
generalized entropy.

\section{Hawking radiation}

Using the techniques of the previous section, we can compute the entanglement entropy of Hawking radiation via the replica trick in gravitational theories without using an auxiliary boundary QFT description. 
We consider situations in which the Hawking radiation propagates to an asymptotic region
where gravity is weak. In the path integral, this means that geometries which deviate substantially
from the semiclassical one in this asymptotic region give a negligible contribution, and therefore the
geometry there can be treated as effectively fixed. We include only perturbative variations, realized
as gravitons propagating on this fixed background. The general idea is to choose a codimension-one
surface $R$ in the region with fixed (or semiclassically controlled) geometry, on which the Hawking
radiation is collected, and to use the gravitational replica trick to compute the entanglement entropy
associated with this region.

\subsection{The replica path integral}

In \ref{Bulk RPI} we saw how to construct a replica path integral in a gravitational region using the
equivalence with the QFT replica path integral. In this section we do not assume a QFT dual to the
gravitational theory, but the procedure used in \ref{Bulk RPI} can be generalized in a purely bulk
language. The key difference is that we now want to calculate the entanglement entropy of a surface
$R$ defined \emph{inside} the gravitational theory. This surface plays a role analogous to $\Sigma_A$
rather than $A$: it is part of the gravitational Cauchy data, and it is fixed by definition (whereas
$\Sigma_A$ in the holographic construction is selected dynamically within the bulk replica saddle).
For clarity, we restate the replica construction in this setting, even though many steps closely parallel those of Section~2 and Section~3.

Consider a gravitational model in which a region exhibits weak gravitational effects. Within this
region, we select a non-timelike codimension-1 surface $R$, described by a density matrix
$\hat{\rho}_R$. We denote by $\Sigma$ a (full) Cauchy slice containing $R$, and by $\bar{R}$ the
complement of $R$ on $\Sigma$, such that $R \cup \bar{R} = \Sigma$. We choose the system on $\Sigma$ to
be in a state $\ket{\Psi}$; for example, we fix the induced metric and field configurations on a Cauchy
slice $\Sigma'$ in the past of $\Sigma$.

Using some version of the no-boundary proposal \cite{Hartle:1983ai}, we can express the wave
functional $\Psi[h^\alpha_{kl},\phi^\alpha_s]$, related to a state
$\ket{h^\alpha_{kl},\phi^\alpha_s}\equiv\ket{\alpha}$ on the Cauchy slice $\Sigma$, as a path integral
over all geometries compatible with the state $\ket{\Psi}$ and with future boundary the slice $\Sigma$.
Here $\phi_s$ represents all non-metric fields of the theory and $h_{kl}$ is the induced metric on the
Cauchy slice $\Sigma$. On the future boundary, the induced metric and all field configurations have
fixed values, $h_{kl}=h^\alpha_{kl}$ and $\phi_s=\phi^\alpha_s$. The wave functional is given by
\begin{gather}
\Psi(\alpha)=\braket{\alpha}{\Psi}
=N^{-\frac{1}{2}}\int_\Psi^{\ g_{kl}=h^\alpha_{kl},\phi_s=\phi^\alpha_s,{\ \rm on\ } \Sigma} [dgd\phi_s]
\, e^{iS[g,\phi_s]},
\end{gather}
where $N^{-1/2}$ is a normalization constant. The past boundary conditions are determined by the state
$\ket{\Psi}$. The path integral is taken over all geometries with finite action and finite field
configurations that satisfy the boundary conditions \cite{Hartle:1983ai}. If the state $\ket{\Psi}$
fixes the values of the induced metric and field configurations on $\Sigma'$, the path integral takes
the form of a propagator from the past Cauchy slice $\Sigma'$ to the future Cauchy slice $\Sigma$. We
include all geometries that match the boundary conditions on the two Cauchy slices.

For the remainder of the section we choose the state on the Cauchy slice to be pure. Such a state is
described by the density matrix $\hat{\rho}=\ket{\Psi}\bra{\Psi}$. Similar to \ref{raa1}, we can use a
basis where the matrix elements of $\hat{\rho}$ are given by two path integrals with boundary
conditions on the surface $\Sigma$,
\begin{gather}
\rho^{\alpha\alpha'}=\bra{\alpha}\hat{\rho}\ket{\alpha'}=\braket{\alpha}{\Psi}\braket{\Psi}{\alpha'}
=\Psi(\alpha)\Psi^*(\alpha').
\label{raa1b}
\end{gather}
To obtain the corresponding matrix elements of the reduced density matrix associated with the
subsystem $R$, we need to sum over the degrees of freedom of the complement $\bar{R}$. Using the
notation $\ket{\alpha}=\ket{\alpha_R}\ket{\alpha_{\bar{R}}}$ and the definition
$\hat{\rho}_R=\Tr_{\bar{R}}\hat{\rho}$, we can write
\begin{align}
\rho_R^{\alpha_R\alpha'_R}&=\bra{\alpha_R}\hat{\rho}_R\ket{\alpha'_R}=
\bra{\alpha_R}\Tr_{\bar{R}}\big(\ket{\Psi}\bra{\Psi}\big)\ket{\alpha'_R}
\nonumber\\
&=\bra{\alpha_R}\bigg(\sum_{\alpha_{\bar{R}}}\bra{\alpha_{\bar{R}}}{\Psi}\rangle\bra{\Psi}\ket{\alpha_{\bar{R}}}\bigg)\ket{\alpha'_R}
=\sum_{\alpha_{\bar{R}}}\braket{\alpha}{\Psi}\braket{\Psi}{\alpha'}
\nonumber\\
&=\int[d\alpha_{\bar{R}}]\Psi(\alpha_R\oplus\alpha_{\bar{R}})\Psi^*(\alpha_R'\oplus\alpha_{\bar{R}})
\nonumber\\
&=\int[d\alpha_{\bar{R}}]
N^{-\frac{1}{2}}\int_\Psi^{\ g_{kl}=h^\alpha_{kl},\ \phi_s=\phi^\alpha_s{\ \rm on\ } \Sigma} [dgd\phi_s] e^{iS_{\uparrow}[g,\phi_s]}
N^{-\frac{1}{2}}\int^\Psi_{\ g_{kl}=h^\alpha_{kl},\ \phi_s=\phi^\alpha_s{\ \rm on\ } \Sigma} [dgd\phi_s] e^{-iS_{\downarrow}[g,\phi_s]}
\nonumber\\
&=N^{-1}\int [dgd\phi_s] \, e^{iS[g,\phi_s]}.
\label{rij1b}
\end{align}
The path integral \ref{rij1b} is over all geometries that match the state $\ket{\Psi}$ and the boundary
conditions $g_{kl}=h^{\alpha_R}_{kl},\phi_s=\phi^{\alpha_R}_s$ on the surface $R$, ``glued''%
\footnote{If the geometry has time-reflection symmetry along the (spatial) $\Sigma$, the path integral
is over geometries with future and past boundaries determined by $\ket{\Psi}$, with a cut along the
region $R$.}
through the region $\bar{R}$ with all geometries that match the state $\ket{\Psi}$ and boundary
conditions $g_{kl}=h^{\alpha_R'}_{kl},\phi_s=\phi^{\alpha_R'}_s$ on $R$. Inspired by the
time-independent case, we can think of the path integral \ref{rij1b} as taken on a bra--ket contour
geometry, with appropriate matching conditions on $\bar{R}$. In the near past, $R^-$, and the near
future, $R^+$, of the cut $R$, we apply boundary conditions%
\footnote{We use the words ``past'' and ``future'' to refer to the regions in the past of $R$ for the
ket and bra geometry, respectively.}.
Note that in the path integral we include geometries where $\bar{R}$ is an arbitrary non-timelike slice
complementary to $R$ on some Cauchy slice. We will call this manifold ${\mathcal{M}}_1$. The
normalization factor $N^{-1}$ is equal to the path integral over ${\mathcal{M}}_1$ without the cut,
ensuring that $\Tr\hat\rho_R=1$.

We are now ready to write the trace $\Tr\hat\rho_R^n$ as a path integral,
\begin{gather}\label{eq:trrn36b}
\Tr\hat\rho_R^n=\sum_{i_1,i_2,\dots,i_n}\rho_R^{i_1 i_2}\rho_R^{i_2 i_3}\dots\rho_R^{i_n i_1}
=N^{-n}\int_{\mathcal{M}_n}[dgd\phi_s]e^{iS[g,\phi_s]},
\end{gather}
where the indices $i_m$ refer to states on the surface $R$. Every matrix element $\rho_R^{ij}$ is a
path integral of the form \ref{rij1b}, and $\Tr\hat\rho_R^n$ involves $n$ copies of the manifold
${\mathcal{M}}_1$. The sum over $i_2$ is equivalent to ``gluing''%
\footnote{As explained in the previous section.} the near-future boundary of $R$ of the first copy with
the near-past boundary of $R$ of the second copy. The same gluing applies to all indices $i_m$. The
last sum (over $i_1$) glues the last copy with the first. We call the resulting manifold
${{\mathcal{M}}}_n$. Note that due to this construction, the full geometry on ${{\mathcal{M}}}_n$ has
(real) conical singularities on the boundaries of $R$. We will see in the next subsection how to treat
these singularities.

If we further include in \ref{eq:trrn36b} all topologies allowed by the boundary conditions of the path
integral, topologies where different copies of the manifold ${\mathcal{M}}_1$ are connected not only
via $R$ can appear. 
As before, we call such connections \textit{replica wormholes}
\cite{Almheiri:2019qdq,Penington:2019kki}.
Note that replica wormholes have two main differences from the connections via $R$. First, they can
connect kets with bras from arbitrary copies, not necessarily all together and in the same order as the
connections via $R$. 
Second, there are no conical singularities at the endpoints of these connections; we assume that the geometry on $\mathcal{M}_n$ is smooth everywhere away from $\partial R$.

\subsection{Island rule from replica wormholes}

The island rule was derived from the existence of replica wormholes for general gravitational models
with time-reflection symmetry in \cite{Almheiri:2019qdq}, building on the replica derivation of
generalized entropy in \cite{Lewkowycz:2013nqa}. In this subsection we adapt this derivation to
arbitrary time-dependent semiclassical geometries by formulating the replica path integral directly in
Lorentzian signature, following the conventions of \cite{Dong:2016hjy}.

Similar to the previous sections, we use the unnormalized replica path integral
\begin{gather}
H_n(R)=\int[dgd\phi_s]e^{iS[g,\phi_s]},
\end{gather}
to express $\Tr\hat{\rho}_R^n$:
\begin{gather}
\Tr\hat{\rho}_R^n = \frac{H_n(R)}{H_1^n}.
\label{san b}
\end{gather}
Equation~\ref{eq:ent_entropy} then takes the form
\begin{gather}
S_R = -\lim_{n \rightarrow 1} \frac{\log H_n(R) - n \log H_1}{n-1}.
\label{svnb}
\end{gather}

We split the action into two parts, a ``gravitational'' part and a matter part,
\begin{align}
S[g,\phi_g,\phi_m]=S_{\rm grav}[g,\phi_g]+S_{\rm matter}[g,\phi_g,\phi_m],
\end{align}
where $\phi_m$ represents the matter fields of the theory and $\phi_g$ represents the remaining
fields, such as the dilaton field in two-dimensional theories. We can integrate out the matter fields,
\begin{align}\label{eq:45b}
H_n(R)=\int[dgd\phi_g]e^{iS_{\rm grav}[g,\phi_g]}
\int[d\phi_m]e^{iS_{\rm matter}[g,\phi_g,\phi_m]}
\equiv\int[dgd\phi_g]e^{iS_{\rm grav}[g,\phi_g]+\log H_{\rm matter}[g,\phi_g]} .
\end{align}
The dominant contributions to \ref{eq:45b} come from saddle points, where $g$ and $\phi_g$ satisfy the
equations of motion associated with the exponent $iS_{\rm grav}[g,\phi_g]+\log H_{\rm matter}[g,\phi_g]$.
The path integral then takes the form
\begin{align}
H_n(R)\sim e^{iS_{\rm grav}[g_c,\phi_{g,c}](\mathcal{M}_n)+\log H_{\rm matter}[g_c,\phi_{g,c}](\mathcal{M}_n)},
\end{align}
where $g_c$ and $\phi_{g,c}$ solve the equations of motion.
Applying this relation to \ref{svnb}, we obtain
\begin{align}\label{4.7b}
S_R=&-\lim_{n\rightarrow1}\frac{iS_{\rm grav}[g_c,\phi_{g,c}](\mathcal{M}_n)-niS_{\rm grav}[g_c,\phi_{g,c}](\mathcal{M}_1)}{n-1}
\nonumber\\
&-\lim_{n\rightarrow1}\frac{\log H_{\rm matter}[g_c,\phi_{g,c}](\mathcal{M}_n)-n\log H_{\rm matter}[g_c,\phi_{g,c}](\mathcal{M}_1)}{n-1}.
\end{align}
As in \ref{4.7}, in the absence of gravity only the second term appears and can be used to define the
entanglement entropy of quantum fields on $R$ (since no replica-wormhole topologies contribute).

From now on, the discussion parallels that of \ref{s gen section}. We assume semiclassical dominance of $\mathbb{Z}_n$-symmetric replica saddles and work near the limit $n\rightarrow1$. 
Under these assumptions, the equations of motion fix the geometry on each of the $n$ segments to agree with the $\mathcal{M}_1$ geometry away from the $\mathbb{Z}_n$ fixed surfaces, with additional regularized conical defects around the codimension-two fixed surfaces $\partial I$ of the $\mathbb{Z}_n$ symmetry. 
Each such fixed surface contributes a term of the form
\begin{gather}\label{eq:cone_contribution b}
i \int dt \, dr \, d^{D-2}x \sqrt{-g}\, R
= \int d\tau \, dr \, d^{D-2}x \sqrt{g}\, R
= 4\pi(1-n)\, \mathrm{Area}(\partial I).
\end{gather}

Note that in this case the endpoints of $R$ are real conical singularities of the replica geometry.
Following the existing literature, we do not include an independent localized contribution to the
gravitational action from these endpoints. However, the matter functional does capture the associated
field-theory entanglement across $R$, yielding the \textit{matter entropy} on $R$
\cite{Calabrese:2004eu}.

\subsection{Using only integer numbers of copies}\label{4.3}

This and the following subsection illustrate a replica-based method that works directly with integer
replica numbers and avoids analytic continuation in $n$. The construction is demonstrated in a very
simple toy model; in more general settings it can be viewed as a formal alternative to analytic
continuation, provided the integer-$n$ replica path integrals $H_n$ are sufficiently well-defined and
under perturbative control.

Up to now, to calculate the entanglement entropy of the system $R$ we used the relation
\begin{align}  \label{sa ch5}
S_R=\lim_{n\rightarrow1}\frac{\log\Tr\hat{\rho}_R^n}{1-n}.
\end{align}
First we calculate $\Tr\hat{\rho}_R^n$ for $n$ near $1$, and then we take the limit $n\rightarrow1$.
Near $n=1$ the analysis simplifies significantly: the geometry on $\mathcal{M}_n$ away from the
$\mathbb{Z}_n$-fixed points is the same as the original geometry on $\mathcal{M}_1$. The only
non-vanishing contributions to \ref{sa ch5} come from the region near the $\mathbb{Z}_n$-fixed points,
which we have shown lead to the island rule. The disadvantage of this method is that it requires analytic continuation to non-integer values of $n$ for quantities that are a priori defined only for integer $n$, such as the replica path integral.
In principle, we can derive the entanglement entropy without analytic continuation in $n$ using the
exact identity \cite{Tomaras:2019sjq}
\footnote{We can obtain the usual form for the entanglement entropy
$S_{\textrm{R}}=-\Tr(\hat{\rho}_R\log\hat{\rho}_R)$ by writing the trace of the density matrix as a sum
over its eigenvalues.}
\begin{align} \label{4.32}
S_{\textrm{R}}=\sum_{n=1}^\infty\sum_{m=0}^n\frac{1}{n}\binom{n}{m}(-1)^m\Tr\hat{\rho}_R^{m+1},
\end{align}
where the traces are given by
\begin{align}\label{eq:trrn2}
    \Tr\hat{\rho}_R^{n}=\frac{H_n}{H_1^n}.
\end{align}

To make the calculation more explicit we use a simple model similar to the one used in
\cite{Penington:2019kki}. We consider the degrees of freedom of the theory to split into two groups:
the degrees of freedom living in the asymptotic region $R$, representing the Hawking radiation, and
their partners living inside a black hole $B$. The black hole has a large number $k$ of possible
internal orthonormal states $\ket{B_i}$. Assuming maximal entanglement of the $B$ degrees of freedom,
the state of the full system is
\begin{align}\label{4.35}
    \ket{\Psi}=\frac{1}{\sqrt{k}}\sum_{i=1}^k\ket{B_i}\ket{R_i},
\end{align}
where $\ket{R_i}$ represents the states of the system $R$.

It is easy to apply the island rule in this model. If we consider no islands, the entropy of $R$ is
simply $\log{k}$. We also consider an island including all the degrees of freedom of the black hole%
\footnote{The island configuration should extremize the generalized entropy. In many studied cases this
is satisfied for an island near the black hole horizon including almost all the black hole degrees of
freedom.}.
The island generalized entropy is the area of the black hole horizon (or the associated gravitational
entropy in two-dimensional theories), which is equal to the coarse-grained entropy $S_{\rm BH}$ of the
black hole. The island formula then takes the form
\begin{align}\label{4.36}
    S_R={\rm min}\big\{\log{k},S_{\rm BH}\big\}.
\end{align}

From \ref{4.35} we obtain the density matrix for the $R$ system,
\begin{align}    \label{4.37}
    \hat{\rho}_R=\Tr_B\hat{\rho}
    =\frac{1}{k}\sum_{i,j=1}^k\ket{R_j}\bra{R_i}\langle B_i|B_j\rangle.
\end{align}
The matrix elements of $\hat{\rho}_R$ are gravitational amplitudes $\langle B_i|B_j\rangle$. In chaotic
black hole systems these amplitudes are expected to behave as random variables, and in the presence of
replica wormholes they are orthogonal up to small corrections of order $e^{-S/2}$, where $S$ is a
characteristic entropy of the system. Following \cite{Penington:2019kki} we write
\begin{align}
    \langle B_i|B_j\rangle=\delta_{ij}+R_{ij},\qquad \sum_iR_{ii}=0.
\end{align}
The amplitudes $\langle R_i|R_j\rangle$ involve the region $R$, where we assume gravitational effects
to be negligible, and therefore $\langle R_i|R_j\rangle=\delta_{ij}$.

To see the connection between $R_{ij}$ and replica wormholes, we apply the above to $\Tr\hat{\rho}_R^2$,
\begin{align}\label{4.39}
    \Tr\hat{\rho}_R^2
    &=\frac{1}{k^2}\sum_{i,j=1}^k\langle B_i|B_j\rangle\langle B_j|B_i\rangle
    \nonumber\\
    &=\frac{1}{k^2}\sum_{i,j=1}^k\left(\delta_{ij}\delta_{ji}+2\delta_{ij}R_{ji}+R_{ij}R_{ji}\right)
    =\frac{1}{k}+\frac{1}{k^2}\sum_{i,j=1}^kR_{ij}R_{ji}.
\end{align}
This trace is calculated from the replica path integral $H_2$. Combining the above relation with
\ref{eq:trrn2}, we get
\begin{align}\label{4.40}
    \frac{1}{k}+\frac{1}{k^2}\sum_{i,j=1}^kR_{ij}R_{ji}=\frac{H_2}{H_1^2}.
\end{align}
There are two main contributions to $H_2$. The first is the no--replica-wormhole saddle, without any
connection between different copies except through $R$. The second contribution comes from the
$\mathbb{Z}_n$-symmetric replica-wormhole saddle where the copies are glued cyclically via $I\cup R$,
with $B\subset I$. Identifying the term $\frac{1}{k^2}\sum_{i,j=1}^kR_{ij}R_{ji}$ with the
replica-wormhole geometry, we interpret the characteristic entropy $S$ as the gravitational entropy
associated with the island%
\footnote{In this approximation we neglect possible finite prefactors in the two terms of order $k^0$.}
\begin{align}
\label{4.41}
    \Tr\hat{\rho}_R^2=\frac{1}{k}+e^{-S_{\rm BH}} .
\end{align}

For general integer $n$, $\Tr\hat{\rho}_R^n$ is dominated either by the no-wormhole geometry, which
contributes a term $1/k^{n-1}$, or by the wormhole geometry connecting all copies through $I\cup R$,
contributing a term $e^{-(n-1)S_{\rm BH}}$ coming from the normalized conical singularity around
$\partial I$ of defect angle $2\pi(1-n)$. Equivalently, at the level of saddle dominance one may write
\begin{align}\label{4.42}
    \Tr\hat{\rho}_R^n \simeq {\rm larger\ of}\left\{\frac{1}{k^{n-1}},e^{-(n-1)S_{\rm BH}}\right\}.
\end{align}
This reproduces the Page-curve behavior: for early times $k\ll e^{S_{\rm BH}}$ the entropy grows as
$\log k$, while for late times $k\gg e^{S_{\rm BH}}$ it saturates at $S_{\rm BH}$.
Applying \ref{4.32}, for $k\ll e^{S_{\rm BH}}$ we get
\begin{align}
S_R
=\sum_{n=1}^\infty\frac{1}{n}\left(1-\frac{1}{k}\right)^n
= \log{k},
\end{align}
and for $k\gg e^{S_{\rm BH}}$ we get
\begin{align}
S_R
=\sum_{n=1}^\infty\frac{1}{n}\left(1-e^{-S_{\rm BH}}\right)^n
=S_{\rm BH},
\end{align}
in agreement with the island rule \ref{4.36}. Note that, as usual in a saddle-point expansion, the
entropy is determined by the dominant saddle: we cannot consistently include both terms in \ref{4.42} at
the same time, since mixed products of $k$ and $e^{-S_{\rm BH}}$ are larger than the subleading
contribution in \ref{4.42}.

\subsection{Non-$\mathbb{Z}_n$-symmetric corrections}

In the previous analysis we assumed that the dominant geometries $\mathcal{M}_n$ in the path integral
have $\mathbb{Z}_n$ symmetry. In this section, we include corrections from geometries without
$\mathbb{Z}_n$ symmetry. To do this we use the simple toy model of the previous section, in which the
$\mathcal{M}_n$ geometries are constructed from $n$ copies of the bra--ket geometry $\mathcal{M}_1$,
with the additional feature that any ket can be connected with any bra via the same surface $I$.
We proceed as follows: first, we compute $\Tr\hat{\rho}^n_R$, allowing for arbitrary connections between copies through the region $I$, including non-$\mathbb{Z}_n$-symmetric configurations. Then, we use \ref{4.35} to compute the gravitational entanglement entropy.

It is convenient to use a different picture from \ref{4.39} to think of $\Tr\hat{\rho}_R^n$.
Starting with the case $n=2$, we write  
\begin{align}
    \Tr\hat{\rho}^2_R=\frac{1}{k^2}\sum_{i_1,j_1,i_2,j_2=1}^k
    \big\{\braket{B_{j_1}}{B_{i_1}}\braket{B_{j_2}}{B_{i_2}}\big\}
    \braket{R_{j_1}}{R_{i_2}}\braket{R_{j_2}}{R_{i_1}} .
\end{align}
The quantity $\big\{\braket{B_{j_1}}{B_{i_1}}\braket{B_{j_2}}{B_{i_2}}\big\}$ has two contributions,
\begin{align}\label{eq:double I braket}
    \big\{\braket{B_{j_1}}{B_{i_1}}\braket{B_{j_2}}{B_{i_2}}\big\}
    =\delta_{j_1i_1}\delta_{j_2i_2}
    +\delta_{j_1i_2}\delta_{j_2i_1}e^{-S_{\rm BH}} .
\end{align}
The first term is the no-wormhole contribution, where each ket is glued to the bra of the same copy.  
The second term corresponds to the configuration in which the ket of the first copy is glued via $I$ to the bra of the second copy, and vice versa. This produces a regularized conical singularity with defect angle $2\pi$, yielding the exponential suppression factor.
We can think of the quantity in braces as follows: we start with two ket and two bra geometries, and glue any ket with any bra through the region $I$. After this gluing, we treat the states $\ket{B}$ as orthonormal.
Combining the above expressions, we find
\begin{align}
    \Tr\hat{\rho}^2_R=\frac{1}{k}+e^{-S_{\rm BH}} .
\end{align}
Note that the extra factor of $k$ in the second term comes from the maximal entanglement between the $B$ degrees of freedom and the Hawking radiation $R$. This feature follows from the form of the state \ref{4.35} and does not hold for a generic state.

The simplest trace that includes non-$\mathbb{Z}_n$-symmetric configurations is
\begin{align}
    \Tr\hat{\rho}^3_R=\frac{1}{k^3}\sum_{i_1,j_1,i_2,j_2,i_3,j_3=1}^k
    \big\{\braket{B_{j_1}}{B_{i_1}}\braket{B_{j_2}}{B_{i_2}}\braket{B_{j_3}}{B_{i_3}}\big\}
    \braket{R_{j_1}}{R_{i_2}}\braket{R_{j_2}}{R_{i_3}}\braket{R_{j_3}}{R_{i_1}} .
\end{align}
Gluing every $\ket{B}$ with any $\bra{B}$, we obtain
\begin{align}\label{4.49}
    &\big\{\braket{B_{j_1}}{B_{i_1}}\braket{B_{j_2}}{B_{i_2}}\braket{B_{j_3}}{B_{i_3}}\big\}=
    \nonumber\\
    &\delta_{j_1i_1}\delta_{j_2i_2}\delta_{j_3i_3}
    + \delta_{j_1i_2}\delta_{j_2i_1}\delta_{j_3i_3}e^{-S_{\rm BH}}
    + \delta_{j_1i_1}\delta_{j_2i_3}\delta_{j_3i_2}e^{-S_{\rm BH}}
    \nonumber\\
    &+ \delta_{j_1i_3}\delta_{j_2i_2}\delta_{j_3i_1}e^{-S_{\rm BH}}
    + \delta_{j_1i_3}\delta_{j_2i_1}\delta_{j_3i_2}e^{-2S_{\rm BH}}
    + \delta_{j_1i_2}\delta_{j_2i_3}\delta_{j_3i_1}e^{-2S_{\rm BH}} .
\end{align}
The factor of two in the exponent arises when all three copies are glued together cyclically, either clockwise or anticlockwise.\footnote{Let us choose the copies to be connected via $R$ in a clockwise manner.}  
The middle terms come from non-$\mathbb{Z}_n$-symmetric configurations and represent partially connected replica geometries, giving subleading corrections to the island formula.
\ref{4.49} then gives
\begin{align}
    \Tr\hat{\rho}^3_R=\frac{1}{k^2}+3\frac{1}{k}e^{-S_{\rm BH}}+\frac{1}{k^2}e^{-2S_{\rm BH}}+e^{-2S_{\rm BH}} .
\end{align}
As in the $n=2$ case, the dominant contribution comes from the first (no-wormhole) term when the number of $B$ states is smaller than $e^{S_{\rm BH}}$, and from the last (fully symmetric wormhole) term when the number of $B$ states exceeds $e^{S_{\rm BH}}$. When $k\sim e^{S_{\rm BH}}$, the middle terms provide finite corrections. Note that the third term is suppressed by a term of order $1/k^2$ compared to the other terms.

To compute the entanglement entropy we need an expression for $\Tr\hat{\rho}^n_R$ at arbitrary $n$. This problem was analyzed in \cite{Penington:2019kki}. Here, thanks to the simplifications of this model, we can compute $\Tr\hat{\rho}^n_R$ explicitly at leading order in $1/k$.

Consider a loop of $m$ copies glued together via $I$. The gravitational contribution from this loop is $e^{-(m-1)S_{\rm BH}}$. We also encounter sums of the form
\begin{align}    
  \sum_{j_l,i_{l+1}=1}^k
  \braket{B_{j_l}}{B_{i_{l+1}}}\braket{R_{j_l}}{R_{i_{l+1}}}=k .
\end{align}
Such a factor appears whenever the bra of the $l$-th copy is glued to the ket of the $(l+1)$-th copy. The dominant contribution from a loop of $m$ copies arises when the copies are glued ``in order'', i.e., the bras are glued with the kets of the next copy, except for the last bra, which is glued with the first ket (and therefore does not give the contribution $k$)\footnote{In \cite{Penington:2019kki} such geometries are called ``planar''.}. 
The total contribution from this loop is $e^{-(m-1)S_{\rm BH}}k^{m-1}$.

The leading contribution to $\Tr\hat{\rho}^n_R$ comes from summing over all configurations of such ordered loops. If $p$ bras are glued to the kets of the next copies, the contribution is
\begin{align}\label{2loops}
    \frac{1}{k^{\,n-1}}\left(ke^{-S_{\rm BH}}\right)^p ,
\end{align}
with $0\leq p\leq n-1$. Summing over $p$ gives\footnote{The corrections remain small even for $n\sim k$, where the $\mathbb{Z}_n$-symmetric geometries dominate.}
\begin{align}\label{eq:trrn corrections}
    k^{n-1}\Tr\hat{\rho}^n_R=
    \left(1+ke^{-S_{\rm BH}}\right)^n
    -(n-1)\left(ke^{-S_{\rm BH}}\right)^{n-1}
    -\left(ke^{-S_{\rm BH}}\right)^n
    +\mathcal{O}\!\left(\frac{1}{k^2}\right).
\end{align}
We now apply \ref{4.32},
\begin{align}\label{corrected sa}
  S_R=&\sum_{n=1}^\infty\sum_{m=0}^n\frac{1}{n}\binom{n}{m}(-1)^m
  \Tr\hat{\rho}_R^{m+1}
  \nonumber\\
  =&\log k
  -\left(1+ke^{-S_{\rm BH}}\right)\log\!\left(1+ke^{-S_{\rm BH}}\right)
  +ke^{-S_{\rm BH}}\log\!\left(ke^{-S_{\rm BH}}\right)
  +\mathcal{O}(1) .
\end{align}
The same result follows from the limit
$S_R=\lim_{n\rightarrow1}\frac{\log\Tr\hat{\rho}_R^n}{1-n}$ applied to \ref{eq:trrn corrections}.
In the regime $k\gg e^{S_{\rm BH}}$ we recover $S_R=S_{\rm BH}$, while for $k\ll e^{S_{\rm BH}}$ we find $S_R=\log k$, in agreement with the island rule. 
The corrections imply a smooth crossover between the no-island and island phase, rather than a sharp transition.
\section{Discussion}

\subsection{Effective theory from a complete local theory}

In \cite{Penington:2019kki}, a factorization problem arising from including replica wormholes in the
gravitational replica path integral is discussed. This issue is clearly visible in the toy model of
\ref{4.3}. In \ref{4.40} and \ref{eq:double I braket}, the first term is the contribution from the
no-wormhole saddle, while the second comes from the $\mathbb{Z}_n$-symmetric wormhole configuration.
The second term implies that the states of the ``gravitational'' subsystem $\ket{B_i}$ cannot be taken
to be exactly orthonormal. In \cite{Penington:2019kki} it is argued that a possible resolution is that
the gravitational state is not pure, but rather an ensemble of states. The exact quantum amplitudes
take the form
\begin{align}
    \langle B_i|B_j\rangle=\delta_{ij}+R_{ij}, \qquad \sum_i R_{ii}=0.
\end{align}
The quantities $R_{ij}$ are of order $e^{-S/2}$, where $S$ is a characteristic (typically large)
entropy of the system.

The factorization problem is further studied in \cite{Hernandez-Cuenca:2024pey}, where it is argued
that for effective theories arising from a complete local theory, factorization of the replica
partition function is incompatible with unitarity of the complete theory. To understand how an
effective theory differs from a complete theory, consider a local complete theory described by the
Euclidean action $I$ on a Riemannian manifold $\mathcal{M}$. We denote by $\phi$ a subset of the quantum
fields of the theory and by $\Phi$ the remaining fields. The theory is described by the partition
function
\begin{align}
    Z=\int_{\mathcal{M}}[d\phi][d\Phi]\,e^{-I[\phi,\Phi]}.
\end{align}
If we are interested only in the physics of the fields $\phi$, we can integrate out the fields $\Phi$,
\begin{align}
    Z_{\phi}=\int_{\mathcal{M}}[d\phi]\int_{\mathcal{M}}[d\Phi]\,e^{-I[\phi,\Phi]}
    \equiv \int_{\mathcal{M}}[d\phi]\,e^{-I_{\Phi}[\phi]},
\end{align}
where $I_{\Phi}[\phi]$ is the effective action for the fields $\phi$. We started from a local theory,
i.e.\ an action $I[\phi,\Phi]$ given by a single integral over $\mathcal{M}$ with finitely many
derivatives, and we end up with an effective action that is, in general, non-local. Integrating out the
$\Phi$ fields leads to an effective action containing multiple integrals that relate the fields $\phi$
at different points on $\mathcal{M}$. In many cases one can still treat the theory as approximately
local, for instance when the non-local terms are suppressed at energies much lower than the
characteristic scale of the $\Phi$ fields. However, in some situations these non-local terms become
important.

The density matrix describing the degrees of freedom of the fields $\phi$ on a surface $A$ is given by
\begin{align}
    \rho_{A,\phi}=\frac{1}{Z_\phi}\int_{\mathcal{M}_1}[d\phi]\,e^{-I_\Phi[\phi]},
\end{align}
where boundary conditions are imposed for the fields $\phi$ on $A^-$ and $A^+$. Here $\mathcal{M}_1$
denotes the manifold $\mathcal{M}$ with a cut along $A$. Note that even if $A$ is a complete Cauchy
slice, non-local terms in $I_\Phi$ connect the two sides of the cut. These connections arise because
the path integral over the $\Phi$ fields involves no cut along $A$.

To compute the information carried by the system $A$, one typically introduces a replica path integral.
Such a path integral involves $n$ copies of the geometry $\mathcal{M}_1$ cyclically glued along $A$. If
the effective action $I_\Phi$ is treated as local, this path integral factorizes when $A$ is a full
Cauchy slice or when $\hat{\rho}_A$ is semiclassically thermal. This factorization corresponds to
integrating a local action on the replica manifold $\mathcal{M}_n$, which is equivalent to integrating
the action separately on each copy $\mathcal{M}_1$. However, this equivalence is spoiled by the
non-local connections in each copy of $\mathcal{M}_1$, originating from the non-local terms in
$I_\Phi$. In \cite{Hernandez-Cuenca:2024pey} it is argued that these connections can lead to the replica
wormholes discussed in the previous sections.

The same picture can be applied when, instead of splitting fields into $\phi$ and $\Phi$, we divide the
degrees of freedom into those of a black hole and those of Hawking radiation. In a holographic setup,
all physical quantities in the gravitational region can, in principle, be computed from the boundary
non-gravitational theory, but the degrees of freedom associated with the black hole and with the Hawking
radiation are typically not localized in strictly disjoint sectors of the boundary theory. This means
that the state of the radiation on the boundary cannot, in general, be obtained by tracing out a simple
spatial region of the complete theory. Such a tracing induces non-local relations between radiation
modes in the boundary theory, analogous to those described above, and can lead to non-factorization of
the replica path integral even when $A$ is a Cauchy slice or when the semiclassical state is thermal.
By holographic duality, this phenomenon should be reproduced in the bulk theory, and replica wormholes
provide a semiclassical realization of this non-factorization. Of course, this does not constitute a
derivation from first principles of the gravitational replica path integral prescription; further
investigation is required.

\subsection{Boundary contributions and constraints in the QES variational problem}

In the saddle-point approximation of the gravitational replica path integral, the dominant contributions
arise from configurations in which $\tilde{A}$ extremizes the generalized entropy. Conceptually, one can
think of integrating out all bulk degrees of freedom in the path integral except for the location of
$\tilde{A}$, so that the effective action governing the remaining integral is the generalized entropy.
However, this variational problem is generally constrained. For example, $\tilde{A}$ cannot deform
beyond the boundary region $A$. Such constraints contribute to the replica saddle in a manner analogous
to bulk stationary points, and therefore the configuration $\tilde{A}=A$ should always be included as a
possible QES
\footnote{This interpretation is meant as a semiclassical and conceptual guide. A general derivation reducing the gravitational replica path integral to an unconstrained variational problem for $\tilde{A}$ is not presently known.}
.

This type of boundary saddle is already captured by the maximin procedure \cite{Wall:2012uf,Faulkner:2013ana,Engelhardt:2014gca}.
In particular, \cite{Engelhardt:2014gca} shows that, for asymptotically AdS spacetimes, the QES
prescription is equivalent to minimizing $S_{\rm gen}(\tilde{A})$ on each bulk Cauchy slice $\Sigma$
containing $A$ and then maximizing over all such slices. On every $\Sigma$, a surface $\tilde{A}$
approaching $A$ is always a possible choice, and no minimum exceeds $\mathrm{Area}(A)$. Consequently,
if no bulk $\tilde{A}$ exists with generalized entropy smaller than $\mathrm{Area}(A)$, the variational
procedure selects the boundary configuration $\tilde{A}\to A$.

Another type of boundary in the variational problem arises in cosmological spacetimes, such as de Sitter
in the context of the bilayer proposal \cite{Franken:2023pni}. In this framework, candidate QESs can be
restricted to lie within specific causal diamonds, which act as dynamical and causal boundaries for the
allowed locations of $\tilde{A}$. Note that surfaces located on the boundary of such causal diamonds
need not have lightlike segments. This situation naturally occurs when $A$ itself has no boundary, for
example when $A$ is one cosmological horizon (screen) and $\bar{A}$ is the other%
\footnote{According to the bilayer proposal, the QFT dual to de Sitter spacetime is defined on the two
cosmological horizons associated with two comoving observers on antipodal trajectories. A bulk Cauchy
slice intersects these horizons in two screens, producing the QFT Cauchy slice.}.
As explained in \cite{Franken:2023pni}, the correct variational problem in this case must include
contributions from $\tilde{A}$ located on the boundaries of the causal diamond. These contributions
again arise as saddle-point configurations reflecting causal and dynamical restrictions on admissible
replica geometries rather than explicit boundary terms in the gravitational action%
\footnote{Here, such ``boundary'' contributions refer to saddle-point effects induced by constraints on
the replica geometries and on the allowed deformations of candidate extremal surfaces. They should be
distinguished from explicit boundary terms added to the gravitational action: the effects discussed
here enter through the structure of the variational problem in the replica path integral, not through
additional localized terms in the action.}.

\subsection{Replica construction for non-subregion subsystems}

Entanglement entropy and other quantum-information quantities can be defined for arbitrary subsystems of
a quantum theory. In the previous sections we focused on subsystems defined by spatial subregions,
containing all degrees of freedom supported in a given region of a QFT or of the bulk spacetime. These
cases are the most extensively studied in the literature and, in holographic settings, admit a geometric
description in terms of extremal or quantum extremal surfaces
\cite{Ryu:2006bv,Hubeny:2007xt,Faulkner:2013ana,Engelhardt:2014gca}.

In principle, however, the replica trick can be applied to subsystems of a much more general nature,
which are not necessarily associated with spatial subregions. Examples include subsets of field species
in a multi-field theory, collections of modes, or more generally subsystems defined by a choice of
subalgebra rather than by spatial localization \cite{Casini:2008wt,Casini:2017roe}. In such cases there
is no natural geometric description of the entanglement entropy.

From the path-integral perspective, the replica construction can nevertheless be implemented by gluing
only the degrees of freedom belonging to the chosen subsystem across replicas. For decoupled sectors,
this procedure leads to independent replica geometries for each sector, with branch points and defects
appearing only for the fields being traced over. This construction is standard in quantum field theory
and does not rely on geometric extremization or on the existence of a codimension-two QES
\cite{Calabrese:2004eu}.

In gravitational and holographic settings, the extension of this construction is far less clear. The HRT
and QES prescriptions are formulated for spatial subregions and rely on the existence of extremal
codimension-two surfaces anchored to the boundary of such regions. For subsystems that are not
associated with spatial subregions, no universal geometric prescription is currently known. From the
replica path-integral viewpoint, this suggests that the sum over replica geometries may be dominated by
saddles that do not admit a simple interpretation in terms of extremal surfaces. Understanding whether
such ``non-subregion'' subsystems admit a consistent gravitational description, and how their
entanglement should be captured semiclassically, remains an open problem. It would be interesting to
apply the Lorentzian replica framework of the previous sections to further investigate the replica path
integrals associated with such subsystems.

A simple illustration is provided by a conformal field theory consisting of $N$ decoupled free scalar
fields. Consider a spatial region $A$ and define a subsystem that includes only $p$ of the scalar
fields restricted to $A$, tracing over both the complement $\bar A$ and the remaining $N-p$ fields.
Since the fields do not interact, the replica path integral factorizes: only the selected $p$ fields
are cyclically glued across the branch cut along $A$, while the remaining fields propagate independently
on each replica sheet. The resulting entanglement entropy is additive and equals $p$ times the entropy
of a single scalar field on $A$. Although well-defined, this entropy does not admit a
description in terms of a single geometric entangling surface. In gravitational settings, analogous
non-geometric replica constructions may require genuinely new saddles, such as bra--ket wormholes
\cite{Chen:2021bra}.

\subsection{Conclusions}

In this work we have presented a unified perspective on Lorentzian (real-time) replica constructions in
time-dependent gravitational settings, synthesizing the construction of \cite{Dong:2016hjy} and \cite{Colin-Ellerin:2020mva,Colin-Ellerin:2021jev} with the
semiclassical replica-wormhole/island framework and with screen-based cosmological setups. By
formulating the replica path integral directly in Lorentzian signature, we avoid relying on Euclidean
continuation or time-reflection symmetry, which is essential for studying dynamical processes such as
black hole formation, evaporation, and cosmological evolution, especially in settings where Euclidean
methods are ambiguous or inapplicable. Our emphasis has been on clarifying the geometric and variational
structures that control Lorentzian replica saddles and on explaining how generalized-entropy and
island-like contributions arise under standard semiclassical assumptions.

Our analysis indicates that the Lorentzian replica framework is not only a computational tool but also a
useful organizing principle for semiclassical quantum gravity: it naturally accommodates constrained
extremization, highlights the role of saddle-point ``boundary'' configurations, and provides a common
language for discussing islands and replica wormholes in time-dependent backgrounds. Within the scope of this
work, the framework can be applied across several settings: holographic QFTs with timelike boundaries,
screen-based cosmological proposals, and intrinsic gravitational subsystems such as Hawking radiation,
where one can treat an exterior region as semiclassically fixed.

For cosmological models such as de Sitter space, the replica construction provides a way to motivate the
generalized-entropy structure used in screen-based proposals (including disconnected surfaces and island
extensions), once a candidate dual description and screen trajectory are assumed. More broadly, the
Lorentzian replica perspective may be useful for probing tensions between semiclassical gravity,
non-factorization of replica path integrals, and the unitarity of an underlying microscopic theory,
although a complete first-principles understanding of the gravitational path integral in these settings
remains an important open direction.

Future work includes the systematic classification of Lorentzian replica saddles and their domains of
validity, a sharper understanding of how constraint-induced boundary configurations emerge directly from
the replica variational problem, and extensions to gauge-invariant and algebraic notions of subsystems.
We hope that the perspective developed here will facilitate further investigations of time-dependent
quantum gravity, helping to relate gravitational path integrals, entanglement structures, and candidate
dual descriptions.

\section*{Acknowledgements}

I would like to thank Nicolaos Toumbas and Fran\c{c}ois Rondeau for useful discussions and comments.



\newpage
\bibliographystyle{jhep}
\bibliography{bib}

\end{document}